\documentclass[]{aastex}

\shorttitle{Orbital Motions of Binaries in Orion South}
\shortauthors{Zapata \& Rodr\'iguez}

\begin{document}

\title{Orbital Motions of Binaries in Orion South}

\author[0000-0003-2343-7937]{Luis A. Zapata}
\affiliation{Instituto de Radioastronom\'{\i}a y Astrof\'{\i}sica\\
Universidad Nacional Aut\'onoma de M\'exico, Apdo. Postal 3-72, Morelia, Michoac\'an 58089, Mexico}
\email[show]{l.zapata@irya.unam.mx} 
\author[0000-0003-2737-5681]{Luis F. Rodr\'{\i}guez}
\affiliation{Instituto de Radioastronom\'{\i}a y Astrof\'{\i}sica\\
Universidad Nacional Aut\'onoma de M\'exico, Apdo. Postal 3-72, Morelia, Michoac\'an 58089, Mexico}
\affiliation{Mesoamerican Center for Theoretical Physics\\
Universidad Aut\'onoma de Chiapas, Tuxtla Guti\'errez, Chiapas 29050, Mexico}
\email[show]{l.rodriguez@irya.unam.mx}

\begin{abstract}
We present high-angular resolution ($\simeq 0\rlap.{''}06$) 
VLA and ALMA observations of Orion South separated by 15.52 years. The purpose of
this study was to search for orbital motions in three close ($\simeq 0\rlap.{''}1$) binary systems in the region.
We do not detect changes in the position angle of the binaries but in two of the cases we detect
significant changes in their separation in the plane of the sky. We use these changes to estimate that the
total mass of the binaries is in the $\simeq$1-2 $M_\odot$ range. We also estimate the disk
masses from the mm emission. The dust-to-stellar mass ratio is in the 
range of 0.04 to 0.18, values consistent with those expected for very early stellar evolution (Class 0) protostars.
\end{abstract}

\keywords{\uat{Binary stars}{154} --- \uat{Orbital motion}{1179} --- \uat{Radio continuum emission}{1340} --- \uat{Stellar masses}{1614}  }


\section{Introduction} \label{sec:intro}

With a bolometric luminosity of \(\sim10^{4}\,L_{\odot}\) and a total gas mass of \(\sim 10^{2} -10^{3}\,M_{\odot}\) \citep{Mezger1990,Zapata2004,Tang2010,Osorio2017}, 
OMC1-S (also known as OrionSouth, or Orion-S) constitutes one of the most luminous and massive condensations in the Orion A cloud. It is still actively forming stars, as its ``twin'', the Becklin-Neugebauer/Kleinmann-Low (BN/KL) region, located toward the northwest of the Trapezium.  Sensitive millimeter and centimeter continuum observations carried out with the VLA and ALMA have revealed a cluster of 
more than 20 compact sources within a region of approximately \(1'\) centered on OMC1-S \citep{Zapata2004,zap2007,aina2018}. 
The nature of these objects appears to be diverse; however, their millimeter emission suggests that some of them are associated with 
compact circumstellar disks with a size of only 50 au and masses of about 0.05 M$_\odot$ \citep{zap2007}. 

The 7 mm VLA observations of the Orion South region revealed two binary systems: 
139$-$409 and 134$-$411\citep{zap2007}. These binary systems are compact, with an angular separation of $\sim0\rlap.{''}1$, with each
stellar component having 
associated a circumstellar disk. Additionally, each binary is surrounded by a
flattened and rotating molecular structure with size of a few hundred astronomical units, 
suggestive of a circumbinary molecular ring.  The dynamical masses derived for the binary systems from the rotation
of the circumbinary rings are
 \(\gtrsim 4\,M_{\odot}\) for 139$-$409 and \(\gtrsim 0.5\,M_{\odot}\) for 134$-$411 \citep{zap2007}. Our reanalysis of
 the 7 mm image revealed a third binary system, 132$-$413, of similar characteristics of the two previously known.
 
In this study, we attempt to estimate the stellar masses of the components forming these three binary systems, 
139$-$409, 134$-$411, and 132$-$413 searching for variations in their angular separation in the sky.
This method relies on the relative proper motions between the components and complements the 
dynamical masses derived from the radial velocities across the circumbinary rings.
By making reasonable assumptions, we derive estimates
for the stellar masses of these binary systems located in Orion~South and discuss the implications.

\section{Observations} \label{sec:obs}

\subsection{VLA 7 mm} 

The observations were conducted with the \textit{NRAO Very Large Array (VLA)} in 
continuum mode at 7~mm (43.34~GHz) on 2004 November~10 and 17,  December 9, and 2005 January 7, as part of project \texttt{AZ154}. 
The data were obtained with an effective bandwidth of 100~MHz. At the time of the observations, 
the array was in its A configuration, providing projected baselines from 0.68 to 36~km (i.e., 99--5200~k$\lambda$). 
The absolute amplitude calibrator was J1331$+$305, with an adopted flux density of 1.45~Jy, while the phase calibrator 
was J0541$-$056, with a bootstrapped flux density of 1.10~Jy. The bandpass calibration was derived from observations of 3C84. 

The data were acquired and reduced following the standard \textit{VLA} procedures for high-frequency observations, including 
the fast-switching mode with a cycle time of 120~s. Imaging was performed using the task \texttt{IMAGR} in \texttt{AIPS}, 
with the \texttt{ROBUST} parameter set to 0. The resulting image achieved an angular resolution of 0\farcs065~$\times$~0\farcs053 
with a position angle of $-7\fdg7$. 
The 7~mm continuum image was also corrected for primary beam attenuation. At this wavelength, the primary beam 
has a full width at half maximum (FWHM) of approximately 60\arcsec. The estimated uncertainty in the absolute flux density 
scale is about 10\%, and the astrometric accuracy is better than 0\farcs02.

These \textit{VLA} data have been previously presented in the millimeter continuum studies of OMC-1S by \citet{zap2007} and \citet{aina2018}.

\subsection{ALMA 3 mm} 

The \textit{Atacama Large Millimeter/Submillimeter Array (ALMA)} Band 3 archive data of the Orion Nebula Cluster (ONC) were carried out 
in Cycle 6 under the project 2018.1.01107.S (PI: Ballering, N.).  The data were acquired on 2019 July 7 and 9 in C43-9 confuration.
In order to cover the central region of the ONC, a small mosaic of ten pointings was made, arranged in a Nyquist-sampled grid. 
The corresponding FWHM at this band is approximately 51\arcsec. The OMC-1S was well covered in this mosaic. 
The total on-source integration time was 198~minutes, corresponding to about 20~minute per mosaic pointing.
The observations employed four spectral windows centered at 90.5, 92.4, 102.5, and 104.4~GHz. Each spectral window had a total bandwidth 
of 1.875~GHz, divided into 1920 channels.  The continuum image was constructed by averaging the line-free channels from the
four spectral windows.
The central frequency for this observation is 97.4375 GHz. 

The observing conditions were generally favorable for this frequency band, with stable atmospheric performance throughout the run. 
The precipitable water vapor averaged about 1.9~mm, and the system temperature was typically 60~K. Standard \textit{ALMA} calibration 
procedures were applied, including the use of water vapor radiometers monitoring the 183~GHz water line to correct for short-term 
atmospheric phase variations. The following quasars served as calibrators: J0423$-$0120 (Amplitude, Atmosphere, Bandpass, Pointing, Water Vapor Radiometer), 
J0529$-$0519 (Gain, Water Vapor Radiometer), J0532$-$0307 (Water Vapor Radiometer), and J0606$-$0724 (Phase, Water Vapor Radiometer). 

Data calibration and imaging were carried out with \texttt{CASA}~5.4.0-70 \citep{mcmullin2007}, using the standard \textit{ALMA} 
calibration pipeline and additional manual processing for imaging and analysis. 
The resulting synthetized beam is $0\farcs071 \times 0\farcs055$ with a position angle of $54\fdg1$.
Briggs weighting was used with a robust parameter of 0.5, in a compromise between sensitivity and angular resolution. 

These \textit{ALMA} data have been previously presented in the millimeter continuum studies of the ONC by \citet{ball2023}.


\begin{deluxetable*}{ccccccccc}\label{tab11}
\tablewidth{0pt}
\tabletypesize{\scriptsize}
\tablecaption{Projects used in this paper \label{tab:description}}
\tablehead{
\colhead{} & \colhead{} & \colhead{Frequency} & \colhead{Number} & \colhead{Mean}  & \colhead{Mean} 
& \colhead{Amplitude}  & \colhead{Gain}  & \colhead{Synthesized}  \\
\colhead{Project} & \colhead{Instrument} & \colhead{(GHz)} & \colhead{of Epochs} & \colhead{Epoch}  & \colhead{MJD}  
& \colhead{Calibrator}  & \colhead{Calibrator} & \colhead{Beam} }
\startdata  
AZ154 & VLA & 43.4 & 4 &  2004-Jan-01 (2004.00)  & 53329.68 & J1331+3030 & J0541$-$0541 
& $0\rlap.{''}065$$\times$$0\rlap.{''}053$; $-7\rlap.^\circ7$ \\
2018.1.01107.S & ALMA & 97.4 & 2 &  2019-Jul-09 (2019.52) & 58670.50 & J0423$-$0120 & J0529$-$0519 
& $0\rlap.{''}071$$\times$$0\rlap.{''}055$; $54\rlap.^\circ1$ \\ 
\enddata

\end{deluxetable*}

\begin{figure*}
\plottwo{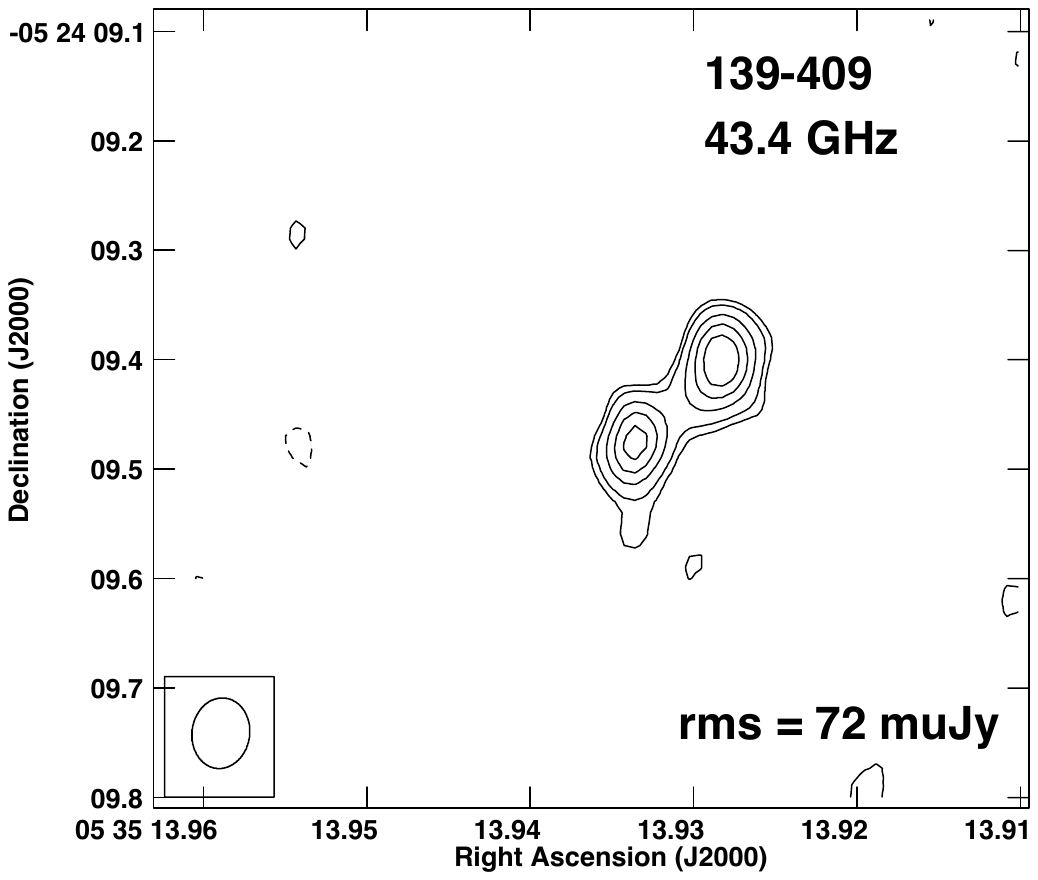}{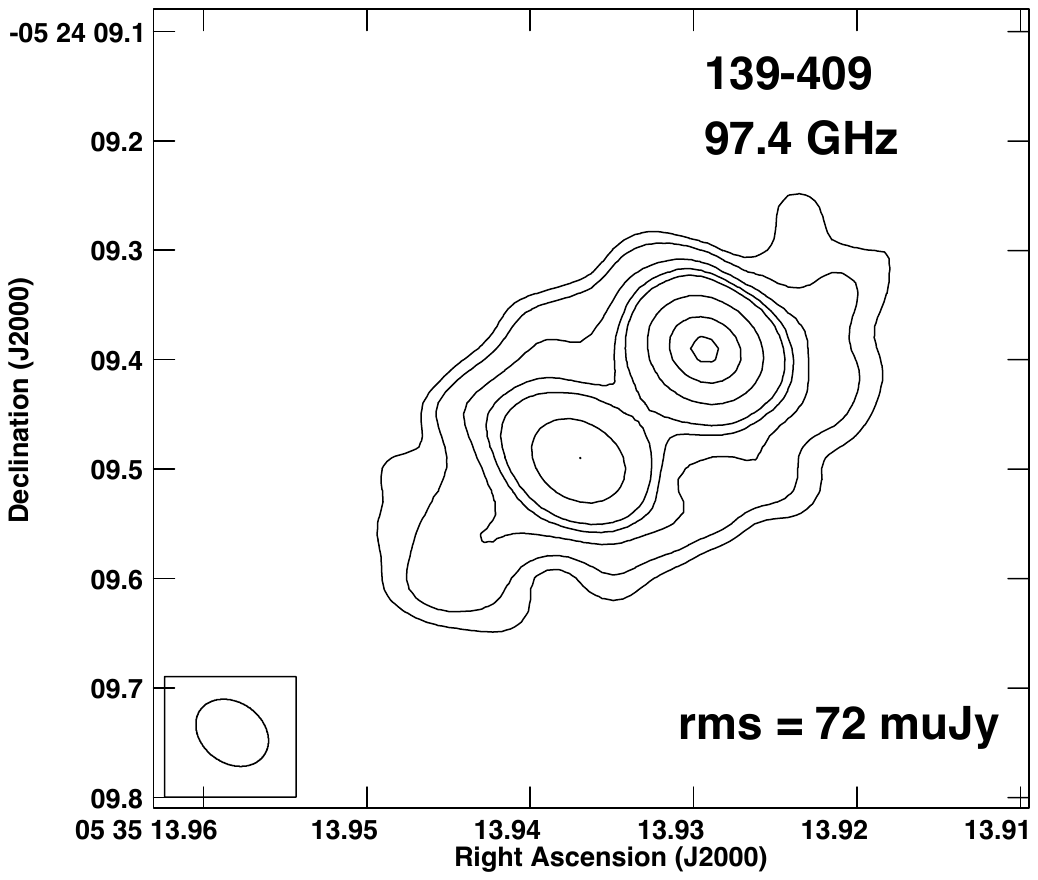}
\vskip-4.5cm
\plottwo{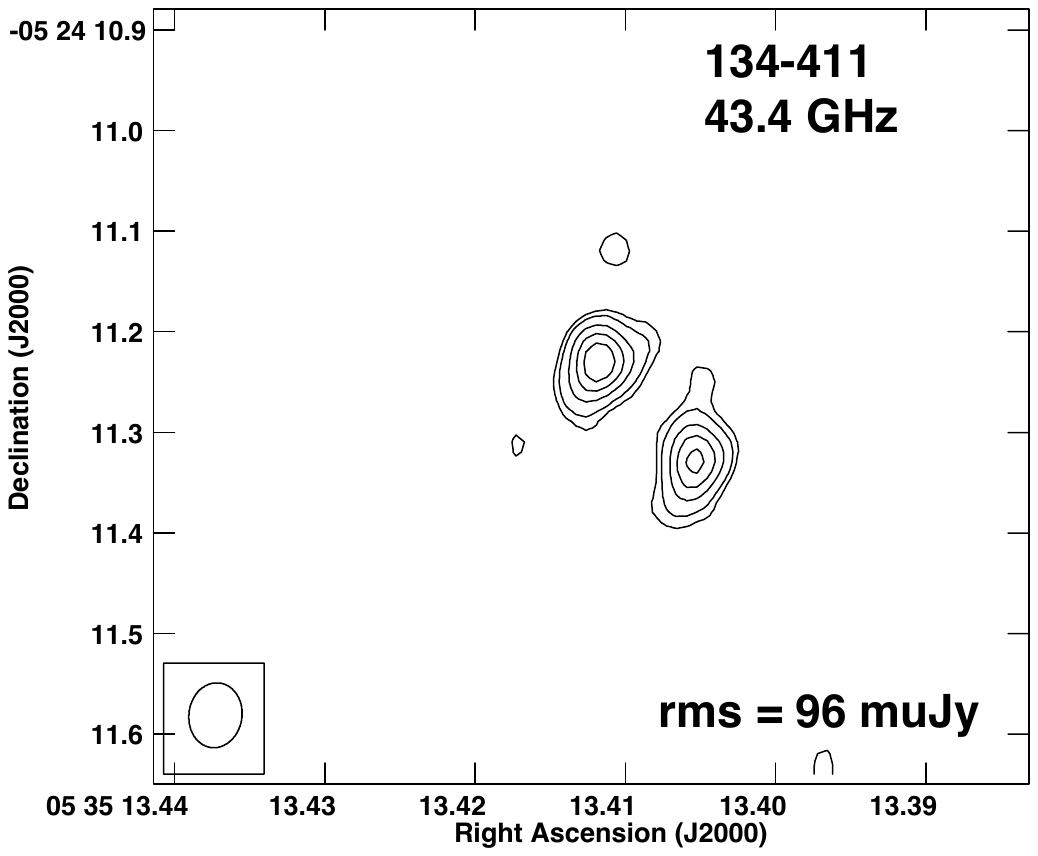}{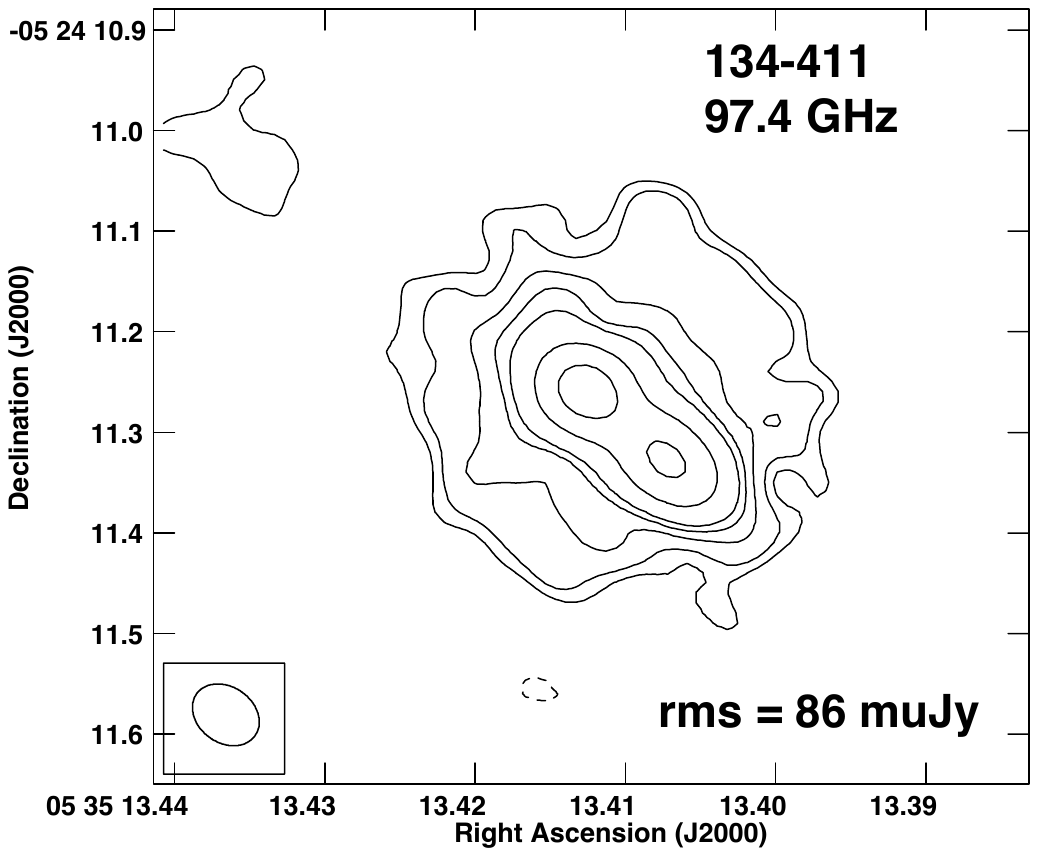}
\vskip-4.5cm
\plottwo{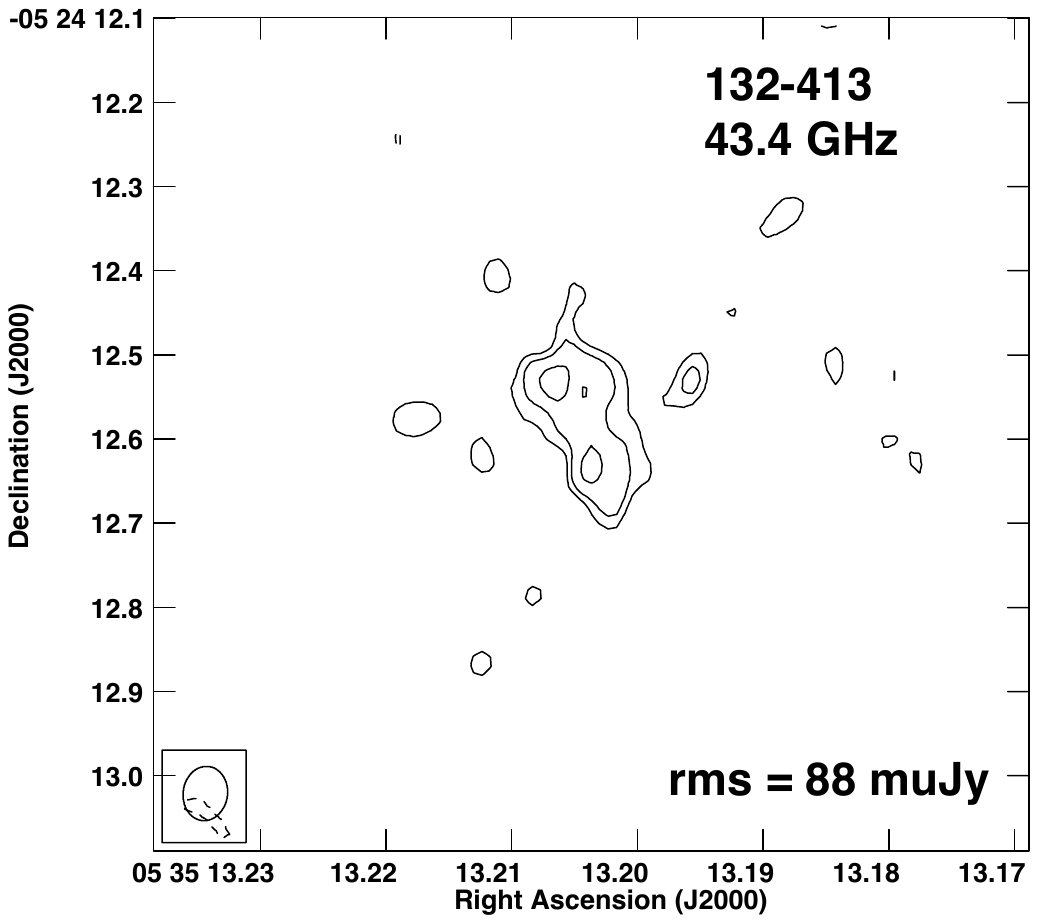}{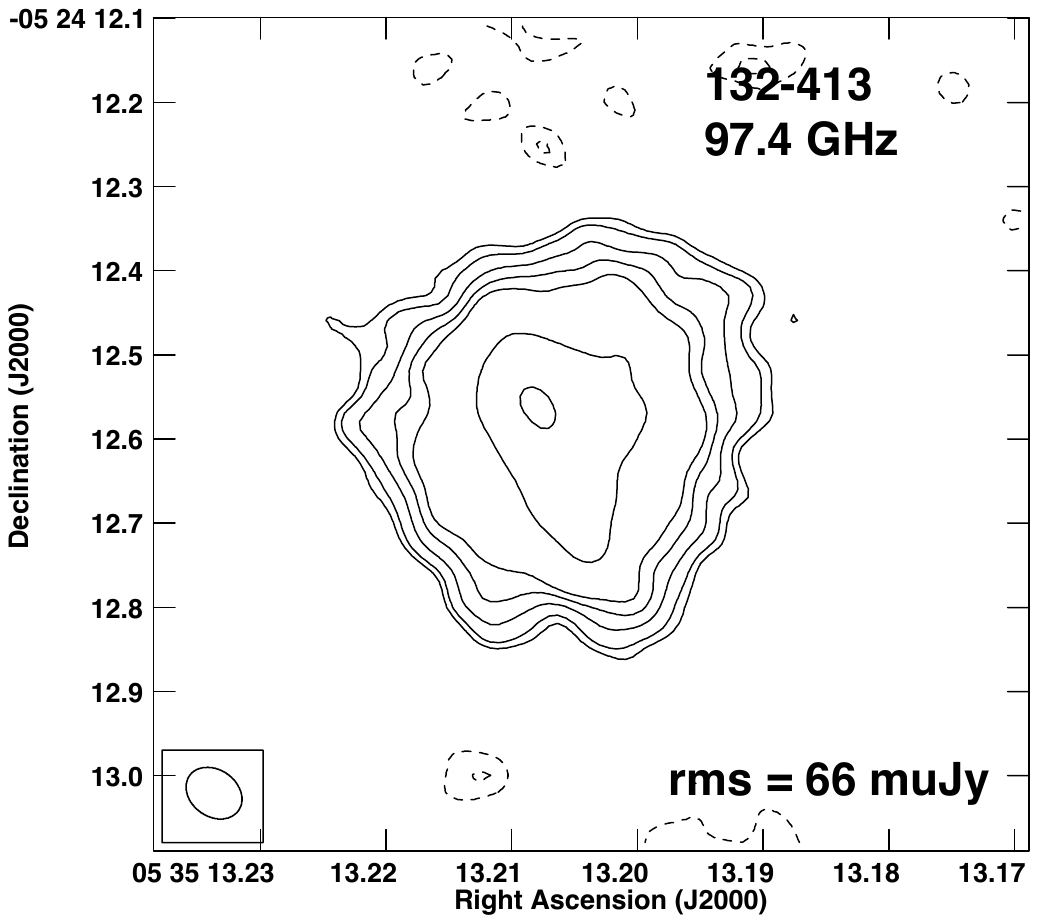}
\vskip-2.0cm
\caption{VLA (left) and ALMA (right) images of 139$-$409 (top), 134$-$411 (middle) and
132$-$413  (bottom). Contours are -4, -3, 3, 4, 6, 8, 10, 20, 40, 60, 80 and 100 times the rms indicated in each
panel. }
\label{fig1}
\end{figure*}

In Table \ref{tab11}  we summarize the parameters of the observations, listing the project name, the instrument used,
the frequencies and the number of epochs observed. We also list the mean epoch and Modified Julian Date (MJD), the
amplitude and gain calibrators and finally the synthesized beam obtained.
The images were corrected for the primary beam response and the binaries are presented in Figure \ref{fig1}.

\section{Analysis}

We fitted the binary sources using simultaneously two Gaussian ellipsoids with the task IMFIT of CASA. We note that
the extent of the sources at 97.4 GHz is larger than at 43.4 GHz. This comes most probably from the fact that at lower
frequencies we detect only the inner part of the disk \citep{2021MNRAS.506.2804T}.  
The positions determined
for both epochs are given in Table \ref{tab2}. In this Table we also give the separation and position angle of the binaries. 
Finally, the last column gives the separation and position angle change between epochs. We first note that the difference in position angle is not
statistically significant and is in the 1 to 2-$\sigma$ range. In contrast, the separation difference is significant at the 7 and 11-$\sigma$ for 
139$-$409 and  134$-$411, respectively. 

Taking into account a distance of 388 pc \citep{2017ApJ...834..142K}, 
the mean separation between the components are 51$\pm$2 and 38$\pm$1 au 
for 139$-$409 and 134$-$411, respectively. Finally, for a time separation of 15.52 years, the proper motions imply plane-of-the-sky
relative velocities of $+$5.0$\pm$0.8 and $-$9.3$\pm$0.8 km s$^{-1}$ for 139$-$409 and 134$-$411, respectively.
The plus (minus) sign represents increasing (decreasing) separation.

In the case of source 132$-$413 we did not detect significant changes in the position angle or the separation.

\begin{deluxetable*}{cccccccc}
\tablenum{2}
\tablecaption{Positions of the Binaries in Orion South\label{table2}}
\tablewidth{900pt}
\tabletypesize{\scriptsize}
\tablehead{
\colhead{} &  
\multicolumn{2}{c}{Position 2004.00 }  &
&
\multicolumn{2}{c}{Position 2019.52 }  &
\colhead{} &
\colhead{} \\
\cline{2-3}
\cline{5-6}
\colhead{Source} & 
\colhead{RA(J2000)}  & 
\colhead{DEC(J2000)} &
\colhead{$\theta$($''$); PA($^\circ$)} &
\colhead{RA(J2000)}  & 
\colhead{DEC(J2000)} &
\colhead{$\theta$($''$); PA($^\circ$)} &
\colhead{$\Delta \theta('')$;
$\Delta$PA($^\circ$)} }
\decimalcolnumbers
\startdata
139$-$409a &
13$\rlap.^s$9282$\pm$0$\rlap.^s$0002 & 09$\rlap.''$403$\pm$0$\rlap.''$004  &
\nodata &
13$\rlap.^s$9293$\pm$0$\rlap.^s$0001 & 09$\rlap.''$391$\pm$0$\rlap.''$001  &
\nodata  &  \nodata \\
139$-$409b &
13$\rlap.^s$9336$\pm$0$\rlap.^s$0003 & 09$\rlap.''$481$\pm$0$\rlap.''$006  &
0.112$\pm$0.007; 133.9 $\pm$3.3 &
13$\rlap.^s$9371$\pm$0$\rlap.^s$0001 & 09$\rlap.''$491$\pm$0$\rlap.''$001  &
0.153$\pm$0.001; 130.8 $\pm$0.4  &  0.041$\pm$0.007; $-$3.1$\pm$3.3 \\
 & & & & & & & \\
 134$-$411a &
13$\rlap.^s$4054$\pm$0$\rlap.^s$0003 & 11$\rlap.''$332$\pm$0$\rlap.''$007  &
\nodata &
13$\rlap.^s$4098$\pm$0$\rlap.^s$0001 & 11$\rlap.''$294$\pm$0$\rlap.''$001  &
\nodata  &  \nodata \\
134$-$411b &
13$\rlap.^s$4117$\pm$0$\rlap.^s$0002 & 11$\rlap.''$232$\pm$0$\rlap.''$005  &
0.137$\pm$0.007; 43.3$\pm$2.9 &
13$\rlap.^s$4129$\pm$0$\rlap.^s$0001 & 11$\rlap.''$256$\pm$0$\rlap.''$001  &
0.060$\pm$0.001; 50.6$\pm$1.3  &  -0.077$\pm$0.007; 7.3$\pm$3.2 \\
& & & & & & & \\
132$-$413a &
13$\rlap.^s$2031$\pm$0$\rlap.^s$0005 & 12$\rlap.''$620$\pm$0$\rlap.''$011  &
\nodata &
13$\rlap.^s$2050$\pm$0$\rlap.^s$0003 & 12$\rlap.''$684$\pm$0$\rlap.''$003  &
\nodata  &  \nodata \\
132$-$413b &
13$\rlap.^s$2055$\pm$0$\rlap.^s$0007 & 12$\rlap.''$549$\pm$0$\rlap.''$013  &
0.080$\pm$0.016; 26.4$\pm$9.8 &
13$\rlap.^s$2060$\pm$0$\rlap.^s$0001 & 12$\rlap.''$607$\pm$0$\rlap.''$002  &
0.079$\pm$0.004; 11.3$\pm$3.1  &  -0.001$\pm$0.016; $-$15.1$\pm$10.3 \\
\enddata
\tablecomments{The positions given show only the seconds of the RA and the arcseconds of the DEC. The remaining values are
$RA(J2000) = 05^h~35^m$ and $DEC(J2000) = -05^\circ~24{'}$.}
\label{tab2}
\end{deluxetable*}

\begin{deluxetable*}{ccccccc}
\tablenum{3}
\tablecaption{Flux Densities and Masses of the Binaries in Orion South\label{table2}}
\tablewidth{900pt}
\tabletypesize{\scriptsize}
\tablehead{
\colhead{} &  
\multicolumn{3}{c}{Flux Density(mJy)}  &
Spectral &
 Stellar &
\colhead{Disk} \\
\cline{2-4}
\colhead{Source} &
 \colhead{6.1 GHz}  &
 \colhead{43.4 GHz}  & 
\colhead{97.4 GHz} &
\colhead{Index} &
\colhead{Mass($M_\odot$)}  & 
\colhead{Mass($M_\odot$)} }
\decimalcolnumbers
\startdata
139$-$409 & 0.110$\pm$0.003 &
3.4$\pm$0.2 & 24.1$\pm$0.1  &
2.4$\pm$0.1 &
0.8  &
0.14   \\
 134$-$411 & 0.456$\pm$0.003 &
6.4$\pm$0.2 & 25.0$\pm$0.2  &
1.7$\pm$0.1 &
1.9  &
0.07   \\
132$-$413 & $\leq$0.025 &
7.4$\pm$0.3  & 36.5$\pm$0.2  &
2.0$\pm$0.1 &
\nodata  &
0.09   \\
\enddata
\tablecomments{The stellar masses in this table are given assuming the statistical correction described in the text.
The flux density at 6.1 GHz is taken from 
\citet{2016ApJ...822...93F}. }
\label{tab3}
\end{deluxetable*}

\section{Discussion}

\subsection{Stellar Mass Estimate}

With only one projected separation and projected relative velocity it is difficult to estimate the masses of the stars that form the binary.
We can obtain a rough estimate making a series of assumptions. We first assume that the stars have the same mass. This is
consistent with the fact that the emission from their disks is comparable. We also assume that the orbit of the binary is very
eccentric and that the apastron is much larger that the observed separation.  Our assumption is justified by the fact that in
a very eccentric orbit there are only small changes in the position angle and most of the change is in the separation
of the binary.  In this case the mass $M$ of each star is given by:

$$M \simeq {{v_{rel}^2 ~S} \over{4 G}},$$
  
\noindent where $v_{rel}$ is the relative velocity, $S$ is the separation, and $G$ is the gravitational constant. In convenient units 
we have that

$$\Biggl[{{M} \over {M_\odot}}\Biggr]  \simeq 2.8 \times 10^{-4}
\Biggl[{{v_{rel}} \over {km~s^{-1}}} \Biggr]^2 ~\Biggl[{{S} \over {au}} \Biggr].$$

From the parameters derived above we find that $M \simeq 0.4~M_\odot$ for 134-409 and  
$M \simeq 0.9~M_\odot$ for 134-411. The true masses are probably
larger because what we are measuring in the plane of the sky is related to the true values by

$$v_{true} = v_{sky}/sin(\phi); S_{true} = S_{sky}/sin(\phi),$$    

\noindent where $\phi$ is the angle between the line of sight and the relative velocity vector. Then, the true mass is related to the mass determined from the projected parameters by

$$M_{true} = M_{sky}/sin^3(\phi).$$

Since the mean value of $\langle sin^3(\phi) \rangle$ = 3$\pi$/16 = 0.589, we expect that the true masses could be about two 
times larger that given above. We conclude that a reasonable estimate for the total stellar masses of these two binaries are
$\simeq$0.8 and $\simeq$1.9 $M_\odot$ for 139$-$409 and 134$-$411, respectively, as listed in Table \ref{tab3}.

In comparison, the assumption of a circular orbit gives mass values a factor of two larger than those of a highly elliptical orbit, namely

$$M \simeq {{v_{rel}^2 ~S} \over{2 G}}.$$

\subsection{Estimation of disk masses from ALMA 3 mm continuum data}

The total (gas + dust) mass associated with the 3~mm continuum emission can be estimated assuming that the emission is optically thin 
and arises from thermal dust radiation. The mass is given by

\begin{equation}
M = \frac{S_\nu \, d^2}{\kappa_\nu \, B_\nu(T_d)} ,
\end{equation}

\noindent where $S_\nu$ is the flux density at frequency $\nu$, $d$ is the distance to the source, $\kappa_\nu$ is the dust opacity per unit (gas + dust) mass, 
and $B_\nu(T_d)$ is the Planck function for a dust temperature $T_d$. At $\nu = 94.7$~GHz (3~mm) and for $\beta = 1.5$,
we adopt $\kappa_{3\mathrm{mm}} = 0.01~\mathrm{cm^2\,g^{-1}}$ for the total (gas+dust) mass, assuming a gas-to-dust ratio of 100 \citep{oss1994}. 
Assuming a distance of d=388$~\mathrm{pc}$, the flux densities obtained in Table 3, and dust temperatures of T$_d$ = 100$~\mathrm{K}$ for the cases of 134$-$411 and 139$-$409 (they are associated with hot molecular gas) and T$_d$ = 50$~\mathrm{K}$ for the case of 132$-$413, we obtain a mass for 132$-$413 of 0.09 M$_\odot$, for 134$-$411 of 0.07 M$_\odot$, and for 139$-$409 of 0.14 M$_\odot$. 
Lower dust temperatures would yield proportionally higher mass estimates. As the disks 132$-413$ and 139$-$409 are associated 
with warm molecular gas \citep{zap2007}, 
we used a dust temperature T$_d$ = 100$~\mathrm{K}$ in these cases.  All these mass estimates are in very good correspondence 
to the ones obtained at 7 mm \citep{zap2007, aina2018} and are comparable to the median value of 0.075 M$_\odot$ determined
by \citet{2018ApJS..238...19T} for Class 0 protostars.   

\subsection{The Nature of the Emission at 7 and 3 mm}

In Table \ref{tab3} we give the total flux density of the binaries at 43.4 GHz (7 mm) and 97.4 GHz (3 mm). The rapid rise of the flux density 
with frequency between 43.4 and 97.4 GHz implies that the 3 mm flux density is dominated by dust emission. 
This is corroborated by the appearance of the sources in 
Figure \ref{fig1}. While at 3 mm we see extended components that trace individual disks, at 7 mm we see compact components that are tracing the
base of ionized outflows or the inner parts of the disks. In Table \ref{tab3} we also give the total flux density at 6.1 GHz, taken from 
\citet{2016ApJ...822...93F}. The rapid rise in flux density also observed
between 6.1 and 43.4 GHz implies that the 43.4 GHz emission
is dominated by dust as well.  Since the flux densities of the dust components of each binary are similar we tentatively propose that the
masses are also similar. 

In conclusion, both the 43.4 and 97.4 GHz emissions are dominated by dust. The difference in the images comes from the well-known
effect that the lower the frequency, the more compact the appearance of the disk \citep{2019ApJ...883...71C}.

\section{Conclusions}

We have made determinations of the stellar and disk masses of three compact binary systems in Orion South. The stellar masses 
where derived from the proper motions of the stars and are
in the 0.8-1.9 M$_\odot$ range The disk masses were determined from
the mm dust emission and are in the 0.07-0.14 M$_\odot$ range. The disk mass/stellar mass ratio is in the range of
0.04 to 0.18, values consistent with those expected for very early stellar evolution (Class 0) protostars.
Future astrometric observations of these binaries will allow a more precise determination of their stellar masses.

\begin{acknowledgments}
This research has made use of the SIMBAD database, operated at CDS, Strasbourg, France. LAZ acknowledges financial support from CONACyT-280775, UNAM-PAPIIT IN110618, and IN112323 grants, México. LFR thanks the financial support
from grant CBF-2025-I-2471 of SECIHTI, México.
\end{acknowledgments}

\begin{contribution}
LAZ came up with the initial research idea and reduced the data.
LFR was responsible for interpreting the data and for writing and submitting the manuscript.




\end{contribution}

%
\facilities{NRAO VLA, ALMA}

\software{astropy \citep{2013A&A...558A..33A,2018AJ....156..123A,2022ApJ...935..167A},  
          Cloudy \citep{2013RMxAA..49..137F}, 
          Source Extractor \citep{1996A&AS..117..393B}
          }


\bibliography{OrionSouthLib}{}
\bibliographystyle{aasjournalv7}



\end{document}